\documentclass[ reprint,
amsmath,amssymb,
aps,
amsart.cls]
{revtex4-2}

\usepackage{dcolumn}
\usepackage{bm}

\usepackage{graphicx}
\usepackage{subfigure}
\usepackage{graphicx,hyperref}
\usepackage[normalem]{ulem}
\expandafter\let\csname equation*\endcsname\relax
\expandafter\let\csname endequation*\endcsname\relax
\usepackage{amsmath}
\usepackage{amssymb}
\usepackage{multirow, makecell, comment} 
\newcommand{\fla}[1]{\begin{flalign}#1\end{flalign}}
\usepackage{braket}
\usepackage{lineno}
\usepackage{lipsum}
\usepackage{xcolor}
\usepackage{soul}
\usepackage{array}
\usepackage{tabularx}

\begin{document}

\preprint{APS/123-QED}

\title{Optical quantum memory on macroscopic coherence}


\author{  S.A. Moiseev$^{1}$}
\email{s.a.moiseev@kazanqc.org}
\author{K.I. Gerasimov$^{1}$,M.M. Minnegaliev$^{1}$,
E.S. Moiseev $^{1}$
}

\affiliation{$^{1}$Kazan Quantum Center, Kazan National Research Technical University n.a. A.N. Tupolev-KAI, 10 K. Marx St., 420111, Kazan, Russia}

\date{\today}

\begin{abstract}
We propose a quantum memory based on the pre-created long-lived macroscopic quantum coherence. It is shown that the proposed approach provides  new physical properties and methods for retrieval of the signal light fields and improvement of the basic parameters of quantum memory. We demonstrate how the pre-created coherence can enable  quantum storage with low quantum noise, programmable and on demand retrieval of signal light fields in atomic ensembles with natural inhomogeneous broadening. The feasibility of implementing this proposal in various crystals doped with rare earth ions, as well as in atomic gases with a Raman transition indicates a new way for the development of optical quantum memory.
\end{abstract}

                  
\maketitle

\textit{Introduction.-} Current investigations demonstrate increasing attention to the development of various quantum memory (QM) devices becoming one of the key elements of practical quantum informatics \cite{Azuma2023,Ye2024}. 
Despite significant progress \cite{Lukin2003,Lvovsky-NatPhot-2009, Hammerer2010,Chaneliere2018,Guo2023,Lei2023,2024-Moiseev-Physics-Uspekhi}, it was not yet possible to achieve practically significant values for all the main parameters in a single QM device. 
This problem stimulates the search for new approaches to creating QM,
this is especially true for optical QM on multi-atomic ensembles interesting due to its possibility of storing a large number of photonic qubits.

Like a quantum computer \cite{Zhong2020}, QM uses quantum coherence carriers, the difficulties in controlling which at the main stages of QM functioning  cause  the problems in its  implementation.
In the conventional scenario \cite{Lukin2003,Lvovsky-NatPhot-2009, Hammerer2010,Chaneliere2018,Guo2023,Lei2023,2024-Moiseev-Physics-Uspekhi}, the  atoms are prepared initially in one of the long-lived quantum states. 
Then a signal pulse is launched to the QM cell, which, being absorbed, is stored  in the excited atomic coherence. 
The retrieval of the signal at a given time from the QM is carried out by subsequent control of the atomic coherence by using additional classical fields.
Characterizing the conventional  QM protocols, we note that the described scenario of coherence functioning is very limited for a macroscopically large number of atoms used.
Here we want to demonstrate that the very large dimension of the Hilbert space of quantum states of an atomic ensemble could provide a rich opportunity not only to increase the QM information capacity, but also to create the quantum coherence in additional freedom degrees of the QM cell, which can be used to expand its functionality and improve  its basic properties.
Thus, the task arises of revealing new, as yet unknown, hidden resources of quantum coherence \cite{Streltsov2017} of  the atomic ensembles, which could provide new  effective  mechanisms of the QM implementation.


Following this way, in this work we propose an optical QM on atomic ensemble based on using spin \textit{pre-created long-lived macroscopic} (PLM-) coherence as an additional controlled quantum resource.
This approach is developed for the photon echo QM and its applicability to other QM protocols is also shown.
The obtained results demonstrate the power of this approach, which leads to the new useful physical properties of QM and gives new effective methods for  retrieval of signal light fields along with improving basic parameters of optical QM.
We also discuss the existing experimental possibilities for implementing the proposed QM and
the  prospects for using pre-created coherence in the implementation of various quantum protocols.

\textit{Photon echo QM protocol on a spin PLM-coherence.-} The development of QM based on spin/photon echo effect is caused, on the one hand, by the unique properties of  spin and photon echoes \cite{Hahn1950,Kurnit1964} to store a gigantic number of classical light pulses,
on the other hand, by the impossibility to apply the conventional echo schemes due to the resulting quantum noise \cite{Chaneliere2018}. 
The photon echo QM protocols, starting with the controlled reverse of inhomogeneous broadening (CRIB) protocol \cite{Moiseev2001}, offer various options for noiseless control of atomic coherence excited by signal fields  \cite{Chaneliere2018,Guo2023,2024-Moiseev-Physics-Uspekhi}, which, however, have not yet led to the QM devices necessary for practice.
Below we study the main physical properties and advantages of photon echo QM based on using spin PLM-coherence. 

\begin{figure}
\includegraphics[width=0.8\linewidth]{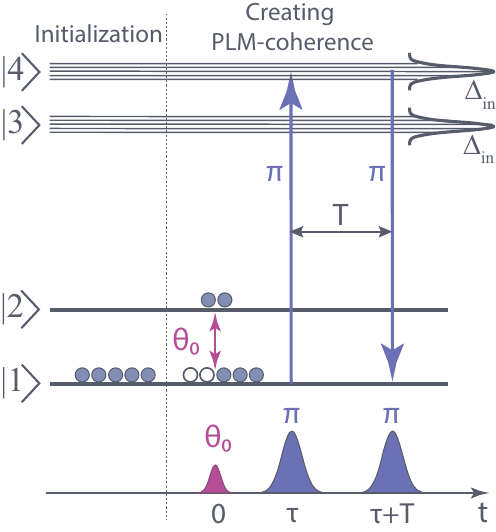}

\caption{\label{Fig:preparing spin coherence}Four atomic levels of the QM with PLM-coherence;  RF pulse with pulse area $\theta_0$ (red arrow) creating spin coherence, two laser  $\pi$-pulses (blue arrows).
Many lines in the states $\ket{3}$, $\ket{4}$  indicate the presence of inhomogeneous broadening ($\Delta_{in}$) at the optical transitions as in other figures.
}
\end{figure}

\textit{A.Basic scheme.-}We use four-level atomic system shown in Fig.\ref{Fig:preparing spin coherence}, which contains two optical levels, each of which is split into 2 sublevels.
For simplicity, we assume that the frequencies of optical transitions are inhomogeneously broadened (with linewidth $\Delta_{in}$), in contrast to the spin transitions of the ground and excited optical levels,  charactrized by long lifetime of spin coherence $T_{2,s}$ at the  transition $\ket{1}\leftrightarrow\ket{2}$. 

Initially we assume that all the atoms are  prepared on the first sublevel 1, 
i.e. $\ket{\Psi_a}_{in}=\prod_{j=1}^{N}\ket{1_j}$, then we apply short   radiofrequency (RF) pulse  
 (with a pulse area $\theta_0$, wave vector  $\textbf{k}_0$ and phase $\varphi_0$) resonant to the spin transition $\ket{1}\leftrightarrow\ket{2}$. 
Exploiting only the basic spin dynamics  within the time scale  $t\ll T_{2,s}$, we leave the consideration of various spin interactions to future research. 
After a small delay $\tau\ll T_{2,s}$, 
we launch a two  laser pulses resonant to the optical transition $\ket{1}\leftrightarrow \ket{4}$,
propagating along $\textbf{k}_1$ and $\textbf{k}_2$  directions with a time delay between them $T\gg T_2^{*}\sim\Delta_{in}^{-1}$ (see Fig.\ref{Fig:preparing spin coherence}) which assumed to be less than the coherence time of the optical transition $T<T_{2,o}$ where usually $T_{2,o}\ll T_{2,s}$. 
We also assume that the signal pulse duration $\delta t_s$ satisfies the condition $T_2^{*}\ll\delta t_s\gg\delta t_{ 1,2}$ so the durations of the laser pulses $\delta t_{ 1,2}$ can be considered infinitely short.
The laser pulses have the same pulse areas $\theta_1=\theta_2=\pi$ and phases  $\varphi_{1}$, $\varphi_{2}$.
After the action of laser pulses the following spin PLM-coherence will be excited:

\fla{
\langle\hat{P}_{12}^j(\tau+T)\rangle
= &
\bra{\Psi_{a}(\tau+T)}\hat{P}_{12}^j
\ket{\Psi_{a}(\tau+T)}
\nonumber \\
=&\frac{i}{2}\sin \theta_0e^{i\Delta_j T}
e^{i\phi+i\delta\textbf{k}_{sc}\textbf{r}_{j}},
\label{P_spin_coherence}
}
where
$\hat{P}_{nm}^j=\ket{n_j}\bra{m_j}$,
$\Delta_j T$ is the phase shift of j-th atom that occurs in free evolution between laser pulses, $\Delta_j$ is a spectral detuning of j-th atom on the optical transition,  $\phi=\varphi_0-\varphi_{1,2}+\omega_{31}T-\omega_{21}\tau$,
$\delta\textbf{k}_{sc}=\textbf{k}_{0}-\textbf{k}_{1}+\textbf{k}_{2}$.

The spin PLM-coherence  \eqref{P_spin_coherence} has the form of a spin wave with the wave vector $\delta\textbf{k}_{sc}$ depended on the wave vectors of the exciting RF- and light pulses.
The spread of frequency offsets ($\{\Delta_j\}$) in Eq. \eqref{P_spin_coherence} 
leads to the suppression of macroscopic coherence and coherent interaction of the spin system with microwave fields, respectively, contributing to the longer storage of PLM-coherence.
Next, we will consider the quantum storage protocols of a light pulse on the prepared atomic ensemble. 






\begin{figure}
\includegraphics[width=1.0\linewidth]{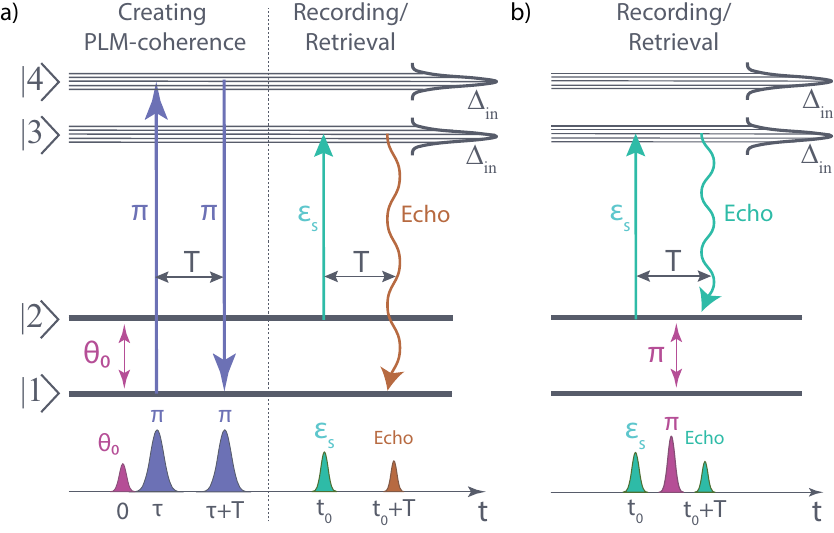}
\caption{\label{Fig:first full scheme} 
Four atomic levels of the QM with PLM-coherence. a) RF-pulse with pulse area $\theta_0$ (red arrow) and two laser  $\pi$-pulses (blue arrows) creates PLM-coherence. Signal pulse and echo pulse are shown by green and yellow wavy arrow, respectively. b) Retrieval of photon echo at the frequency of the input signal. In this case creating PLM-coherence requires RF-pulse with pulse area $\theta_0$=$\pi/2$ and additional RF-pulse with a pulse area of $\pi$ is applied after signal pulse absorption. 
}
\end{figure}

\textit{B.Signal pulse recording stage.-}
The signal light pulse, resonant to the atomic transition $\ket{2}\leftrightarrow \ket{3}$ ($\omega_s=\omega_{32}$), is launched  along z-direction with time delay $t_0-\tau-T$ after the second laser pulse (Fig.\ref{Fig:first full scheme}).
The delay $t_0$ must be shorter than the spin coherence time $T_{2,s}$, but can be longer than the optical transition lifetime $T_{1,o}$,
We describe the quantum light storage  by using Heisenberg equations for the slowly varying operators of signal field $\hat{b}_s(t,z)$ and atomic coherences $\hat{S}^{j}_{nm}(t)$ on the atomic transitions $\ket{n}_j\leftrightarrow \ket{m}_j$
\cite{Scully1997}, taking the initial quantum state of atoms and light at $t=\tau+T$ and
assuming that the initial quantum state of the signal pulse 
$\ket{\psi_{in}(t\rightarrow-\infty)}$
contains, on average, a number of photons much less than the number of atoms in a QM cell.
The solutions have the form 
(see Supplement materials \cite{SM}):



\fla{
\hat{b}_s(t,z) = &
\hat{b}_{s,0}(t-t_s-z/v_g) e^{-\alpha_s z/2},
\label{signal_absorption}
\\
\hat{S}^{j}_{23}(t) = &-ig_s e^{-i\Delta_j(t-t_s)-\alpha_s z_j/2}\sin^2(\frac{\theta_0}{2})\tilde{b}_{s,0}(\Delta_j),
\label{excite_coherence_23}
 \\
\hat{S}^{j}_{13}(t) = &\frac{g_s}{2} e^{-i\Delta_j(t-t_s-T)+
i\phi}
\sin\theta_0 \cdot
\nonumber \\
&e^{i\delta\textbf{k}_{sc}\textbf{r}_{j}
-\alpha_s z_j/2}
\tilde{b}_{s,0}(\Delta_j).
\label{excited_coherence_13}
}
The Eqs. \eqref{signal_absorption}-\eqref{excited_coherence_13} contain only operators describing the properties of the input signal field and atomic coherences excited,
where  $\tilde{b}_{s,0}(\Delta_j) =\int_{-\infty}^{\infty} dt e^{i\Delta_j t}\hat{b}_{s,0}(t)$ is a Fourier component of the input signal,
$\alpha_s=\alpha_{0,s}\sin^2(\frac{\theta_0}{2})$, 
$\alpha_{0,s}=\frac{4\pi N |g_s|^2}{v_g\Delta_{in}L}$ 
is the resonant absorption coefficient on the atomic transition $\ket{2}\leftrightarrow \ket{3}$.

Eq.
\eqref{signal_absorption} shows a complete absorption of an input signal field if  $\alpha_s L\gg 1$ when its quantum state  is transferred into the atomic system with the emergence of two optical coherences $\hat{S}^{j}_{23}(t)$ and $\hat{S}^{j}_{13}(t)$. 
We note that existing photon echo QM protocols (CRIB/GEM, AFC, ROSE,...) \cite{Chaneliere2018,Tittel2010,Guo2023,Lei2023,2024-Moiseev-Physics-Uspekhi} use the signal pulse retrieval  by rephasing the  atomic
coherence  $\hat{S}^{j}_{23}(t)$ \eqref{excite_coherence_23}.
The appearance of an additional optical coherence $\hat{S}^{j}_{13}(t)$ in the studied QM is due to the presence of spin PLM-coherence.
A remarkable property of the coherence $\hat{S}^{j}_{13}(t)$ is that, when excited in a dephased state, it is then macroscopically restored after absorption of the signal pulse with a time delay $T> \delta t_s$.
This can lead to the photon echo emission along the wave vector $\textbf{k}_e (\omega_{31})$  with a carrier frequency $\omega_e=\omega_{31}$  when the phase matching condition is met 

\fla{
\textbf{k}_e (\omega_{31})
=\textbf{k}_{s}+
\delta\textbf{k}_{sc}.
\label{wave_sinh}
}

We study the echo emission in the backward spatial geometry when $\textbf{k}_e (\omega_{31})=-k_e (\omega_{31}) \textbf{e}_z$, providing the  retrieval of signal fields with highest efficiency \cite{Moiseev2001,Gorshkov2007b}.
The backward  geometry is realized here for the following wave vectors of the signal, RF- and two control laser pulses: $ \textbf{k}_{s}\uparrow \uparrow \textbf{k}_{0} \uparrow \uparrow \textbf{k}_{1}\uparrow \downarrow\textbf{k}_{2}$, where $k_0\ll k_{1,2}$ and $k_s<k_{1,2}$.

.

\textit{C. Backward scheme of  echo retrieval.-} We use the equations for slowly varying operators of atomic coherence  $\hat{S}_{13}^j$ and light field $\hat{b}_e(t,z)$, which describe the   echo signal generation in the backward  direction. 
Performing the calculations similar to the absorption stage but taking into account the initial optical coherence $\hat{S}_{13}^j$ \eqref{excited_coherence_13}, we get a solution for the field operator of photon echo emitted (see Supplement materials \cite{SM}):


\fla{
\hat{b}_e(z,t)=&-i\frac{2(\frac{g_s}{g_e})\tan(\frac{\theta_0}{2})e^{i\phi}}{1+|\frac{g_s}{g_e}|^2\tan^2(\frac{\theta_0}{2})}
\hat{b}_s(t-t_s-T+z/v_g)
\cdot
\nonumber \\
&e^{\alpha_e z/2}
\big[
e^{-(\alpha_e+\alpha_s) z/2}-
e^{-(\alpha_e+\alpha_s) L/2}
\big],
\label{Solution_echo}
}
\noindent
where $\alpha_{e}=\alpha_{0,e}\cos^2(\frac{\theta_0}{2})$, $\alpha_{0,e}=\frac{4\pi N |g_e|^2}{v_g\Delta_{in}L}$ is an absorption coefficient on the atomic transition $\ket{1}\leftrightarrow \ket{3}$.

Assuming the atomic medium to be optically depth $(\alpha_e+\alpha_s) L\gg 1$), we obtain at its output

\fla{
\hat{b}_e(z<0,t)=-ie^{i\phi}\hat{b}_s(t-t_s-T+z/v_g),
\label{solution_final}
}
if the condition  
\fla{
|\frac{g_s}{g_e}|\tan(\frac{\theta_0}{2})=1,
\label{symmetry}
}
is satisfied, which means 
equality of the absorption coefficients at both atomic transitions $\alpha_s=\alpha_e$.

The solution  \eqref{solution_final} describes a perfect recovery of the input signal pulse.
Mathematically, this is ensured by symmetry of the used equations written in the variables $\{g_s S_{23}^j, b_s(t,z)\}$ and $\{g_e S_{13}^j, b_e(t,z)\}$ for the  storage and retrieval stages  when the condition \eqref{symmetry} is met together with the spatial inversion $z\rightarrow -z$ in the equations of the photon echo emission.
The found reversibility in the studied optical QM generalizes the reversibility protocols of conventional effective QM schemes \cite{Moiseev2001,Kraus2006,Gorshkov2007,ESMoiseev_2013} to the interactions using different coupling constants ($g_s$ and $g_e$).
Moreover, another carrier frequency of the photon echo ($\omega_e\neq\omega_s$) shows the atomic system  does not return to its initial  quantum state, which distinguishes the physical mechanism of this QM operation from the used in the conventional QM protocols \cite{2024-Moiseev-Physics-Uspekhi,Lvovsky-NatPhot-2009,Chaneliere2018}.
This property is provided by the echo emission in the presence of spin PLM-coherence working as a quantum bath, insensitive to small losses of spin excitations, which allows achieving high efficiency of QM.

\textit{Possible new QM protocols.-}
The studied QM protocol can be modified in various ways, some of them are described below.
For example, we can change the carrier frequency of the echo signal.
 To do this, an additional RF $\pi$-pulse resonant to the spin transition $\ket{1}\leftrightarrow \ket{2}$ should be applied before the echo signal emission. 
 In this case we have to use $\theta_0=\pi/2$ (see Fig.\ref{Fig:first full scheme} b).

\begin{figure}
\includegraphics[width=1\linewidth]{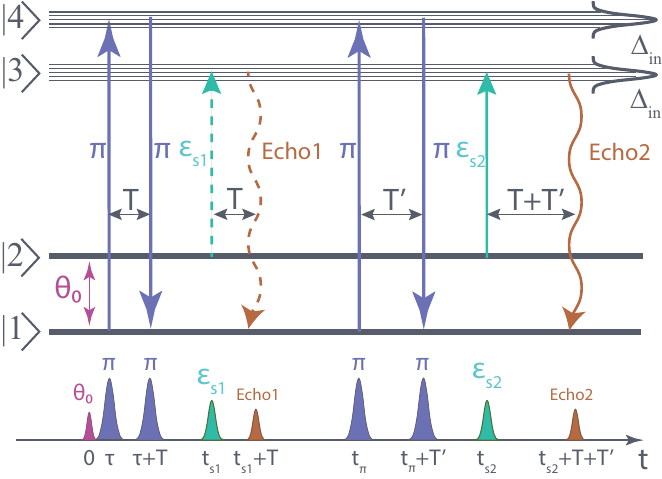}
\caption{\label{Fig:reprogramming increased}Reprogramming with increased storage time of an input signal pulse ($T\rightarrow T+T'$).}
\end{figure}

The  storage time can be dynamically changed by applying two additional  laser $\pi$-pulses before the launching a signal field. 
The two  pulses with similar parameters separated by a time interval $T'$ will increase the previous storage time from $T$ to $T+T'$ (see Fig.\ref{Fig:reprogramming increased}). 
It is also possible to reduce the  storage time from $T$ to $T-T'$  if these additional two laser pulses will acts  at atomic transition $\ket{2}\leftrightarrow \ket{4}$.

\begin{figure}
\includegraphics[width=1\linewidth]{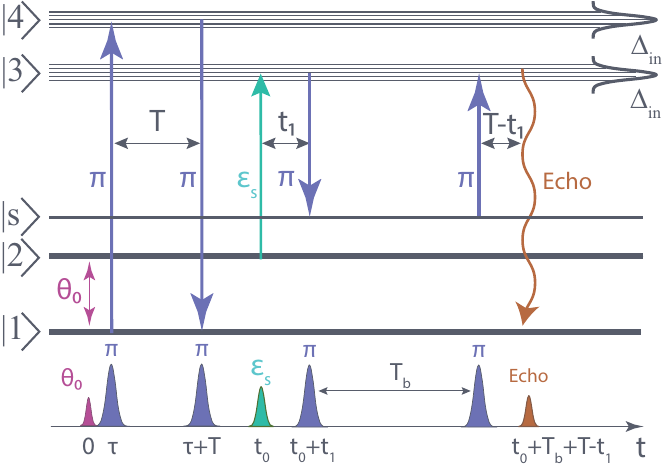}
\caption{\label{Fig:On demand}On demand quantum storage.}
\end{figure}

On demand signal retrieval  is implemented if one  more depopulated spin level $\ket{b}$ is prepared (see Fig.\ref{Fig:On demand}). 
After absorption of a signal pulse, the excited coherence $S_{23}$ should be transferred to the spin coherence $S_{2s}$ by applying a laser $\pi$-pulse at the atomic transition $\ket{3}\leftrightarrow \ket{s}$ with time delay $t_1<T_{2,a}$.
Retrieval of a signal pulse is achieved  similarly to the spin-wave AFC protocol \cite{Afzelius_Kroll2010},  by using a readout laser $\pi$-pulse, resulting in the echo emission   after a delay time $T-t_1$.



A dynamical decoupling technique are often used for increasing the spin coherence time of QMs \cite{2013-Heinze-PRL,2015-Zhong-Nature, 2022-Ortu-npjQI}.
Application of this technique to the spin PLM-coherence will not lead to appearance of quantum noise  due to the using of macroscopic properties of the spin PLM-coherence.


Using cavity-assisted scheme \cite{Moiseev2010cavity,Afzelius-PRA-2010,2013-Sabooni-PRL}, it is possible to reduce the requirements for the optical density of the atomic transitions. 
Moreover, the performing phase matching is also simplified. 
For example, control laser pulses can propagate parallel to each other, being orthogonal to the cavity axis \cite{2021-PRB-Minnegaliev}. 


\textit{Comparison with conventional photon echo QM protocols.-} The proposed optical QM provides a number of advantages.
Comparing the basic properties of the CRIB/GEM-protocols \cite{Moiseev2001,Hosseini2011,2010-Nature-Hedges}, the proposed QM can use full width of natural inhomogeneous broadening, which makes it possible to reduce the concentration of atoms and use a longer lifetime of optical quantum coherence.

Comparing with  AFC protocol \cite{deRiedmatten2008}, instead of creating a periodic classical structure of resonance lines (AFC-structure), in the proposed QM we create a spin PLM-coherence storing  the frequency ($2\pi/T$) and spatial  ($2\pi/|\delta\textbf{k}_{sc}|$) periods (see Eq.\eqref{P_spin_coherence})  
without reducing the optical density of the atomic transition and allowing the use of atomic transitions with  natural inhomogeneous  broadening.
Spin PLM-coherence automatically ensures both the echo emission  at a given moment in time and the reprogramming a storage time during the QM operation.
Thus, spin PLM-coherence can be considered as a quantum generalization of the AFC-structure.

Comparing the proposed QM with the ROSE protocol \cite{2011-NJP-Damon}, we note firstly that it is characterized by the different phase matching condition \eqref{wave_sinh}, which provides a convenient opportunity to retrieve an echo signal in the backward direction.
Apart from the possibility of programmable storage time, the most important advantage is a strongly suppressed level of quantum noise even in comparison with a 4-level ROSE-protocol \cite{Ma2021NLPE}.
Namely, the spin PLM-coherence allows using a time delay for the input signal after the action of two laser $\pi$-pulses that significantly exceeds the lifetime of the optical level.
Such time delay ensures almost complete suppression of the luminescence of atoms excited by the non-ideal controlled  laser $\pi$-pulses and the optical quantum noise caused by this luminescence, respectively.
Moreover, with a large time delay it becomes possible to use one of the two optically excited levels
in Figs.\ref{Fig:preparing spin coherence}, \ref{Fig:first full scheme}, which, with a small delay, make it possible to avoid the appearance of quantum noise in the retrieved light signals due to a considerable change in their frequencies.



\begin{figure}
\includegraphics[width=1\linewidth]{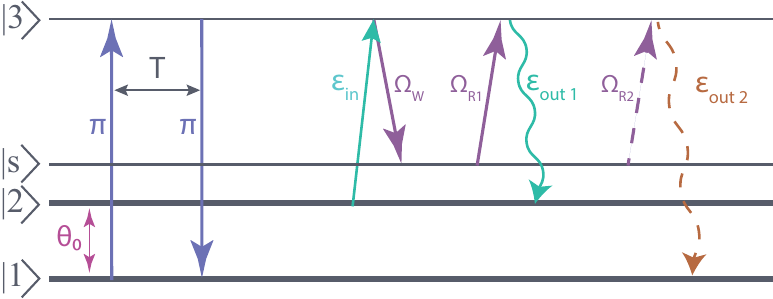}
\caption{\label{Fig:Vapor} Possible use of the spin PLM-coherence in a Raman QM scheme.
}
\end{figure}

\textit{Implementation on Raman transitions.-}
We briefly discuss the possibilities of implementing Raman  QM with $\lambda$-scheme  of atomic transitions, arising due to the use of spin PLM-coherence.
Fig. \ref{Fig:Vapor} shows a schematic diagram of a such Raman QM protocol.
The input signal pulse $\mathcal{E}_s$ is stored 
in the presence of the control laser pulse $\Omega_w$. 
\cite{Lukin2003,Gorshkov2007b,Lvovsky-NatPhot-2009,Radnaev2010}.
The interaction of an input signal in the presence of spin PLM-coherence $S_{12}(\textbf{r})$  leads to the appearance of two spin coherences $S_{1s}$ and $S_{1s}$.

Retrieval of the stored pulse in the output field $\mathcal{E}_{out,1}$ using coherence $S_{2s}$ corresponds to the well-known Raman QM protocols characterized by the (first) phase matching conditions $\textbf{k}_{out,1}=\textbf{k}_{s}-\textbf{k}_{W}+\textbf{k}_{R1}$ \cite{Lukin2003,Lvovsky-NatPhot-2009}. 
In the second possible case, the recovered signal pulse $\mathcal{E}_{out,2}$ will occur due to the presence of coherence $S_{1s}(\textbf{r})$ and irradiate at the changed frequency when the (second) phase matching condition is satisfied: 


\fla{
\textbf{k}_{out,2}=\textbf{k}_{s}-\textbf{k}_{W}+\textbf{k}_{R1}+\delta \textbf{k}_{sc}.
\label{phase_mathc_2}
}
\noindent



Using these phase-matching conditions, it is possible to ensure either the emission of one of the signals ($\mathcal{E}_{out,1}$ or $\mathcal{E}_{out,2}$) with the same high efficiency, or simultaneous emission of the both light fields.
Thus, the presence of the PLM-coherence $S_{12}$ in Raman QM opens up new possibility in creating new  light states.
The considered  QM can be similarly applied to off-resonant Raman quantum transitions.

\textit{Atomic systems promising for experimental implementation.-}
Crystals doped with rare-earth ions are actively used in the development of QMs  \cite{Guo2023,Tittel2010,Ham1997,Turukhin2001}. 
For the proposed QM, it is critical to ensure a sufficiently long spin coherence lifetime. 
Among such crystals a Y$_2$SiO$_5$ host matrix doped with  non-Kramers ions (Eu$^{3+}$ and Pr$^{3+}$)  should be noted. 
These ions have a nuclear spin I=5/2, the states of which are split into three doublets due to the quadrupole interaction in the absence of a magnetic field.
The long lifetime of these states for Eu$^{3+}$ \cite{2003-Konz-PRB} (more than 20 days) and about 100 s for Pr$^{3+}$ \cite{1993-Holliday-PRB}) allows the proposed four-level scheme with high magnetization to be prepared using well-developed optical pumping methods \cite{Nilsson2004}.
The spectral width of a such QM will be limited by the splitting of sublevels due to quadrupole interaction. 
From this point of view, $^{153}$Eu$^{3+}$ (site 1 and 2) and Pr$^{3+}$ (site 1) are preferred. 
The coherence times of the hyperfine and optical transitions (T$_{2HF}$ and T$_{2opt}$) are quite sufficient for the implementation of the proposed QM. 
The main characteristics of Eu$^{3+}$ and Pr$^{3+}$ ions in Y$_2$SiO$_5$ crystal are given in the Table \ref{tab:my_label}.
The achieved ($\approx$ 95$\%$) spin polarization between the hyperfine states $\ket{1}\leftrightarrow \ket{2}$ of the Eu$^{3+}$ and Pr$^{3+}$ ions will only limit the maximum efficiency of the QM, without leading to the optical quantum noises.

\begin{table*}
\caption{The main characteristics of Eu$^{3+}$ and Pr$^{3+}$ ions in Y$_2$SiO$_5$ crystal. Numbers in MHz are splittings of three doublets due to quadrupole interaction of the lowest state in the corresponding multiplet $^{2S+1}$L$_{J}$ where S, L and J are the quantum numbers of total spin, orbital and resultant angular momenta, respectively ($^7$F$_0$ and etc). B is magnetic field.}
\label{tab:my_label}
    \begin{tabular}    {|p{1.0cm}|p{2.4cm}|p{2.6cm}|p{2.4cm}|p{2.6cm}|p{1.0cm}|p{2.4cm}|p{2.0cm}|} \hline  
         \textbf{Eu$^{3+}$} & $^{151}$Eu$^{3+}$, S1 & $^{153}$Eu$^{3+}$, S1 & $^{151}$Eu$^{3+}$, S2 & $^{153}$Eu$^{3+}$, S2 & \textbf{Pr$^{3+}$} & S1 & S2 \\ \hline  
         $^7$F$_0$& 34.533 MHz \cite{2003-Konz-PRB}& 90.0 MHz \cite{1984-Yano-OptLett} & 29.527 MHz \cite{2003-Konz-PRB} & 76.4 MHz \cite{1984-Yano-OptLett} & $^3$H$_4$ & 17.3 MHz \cite{1995-Equall-PRB}& 4.93 MHz \cite{1995-Equall-PRB} \\ \hline  
         &46.175 MHz \cite{2003-Konz-PRB} & 119.2 MHz \cite{1984-Yano-OptLett} & 57.254 MHz \cite{2003-Konz-PRB} & 148.1 MHz \cite{1984-Yano-OptLett} &  & 10.19 MHz \cite{1995-Equall-PRB} & 3.78 MHz\cite{1995-Equall-PRB} \\ \hline  
         $^5$D$_0$& 75 MHz \cite{1984-Yano-OptLett} & 191 MHz \cite{1984-Yano-OptLett} & 63 MHz \cite{1984-Yano-OptLett} & 160 MHz \cite{1984-Yano-OptLett} & $^1$D$_2$ & 4.84 MHz \cite{1995-Equall-PRB} & 2.29 MHz \cite{1995-Equall-PRB} \\ \hline  
         & 102 MHz \cite{1984-Yano-OptLett} & 260 MHz \cite{1984-Yano-OptLett} & 108 MHz \cite{1984-Yano-OptLett} & 274 MHz \cite{1984-Yano-OptLett} &  & 4.59 MHz \cite{1995-Equall-PRB} & 2.29 MHz \cite{1995-Equall-PRB} \\ \hline  
         T$_{1,opt}$&  \multicolumn{2}{|c|}{1.9 ms \cite{1994-Equall-PRL}} &  \multicolumn{2}{|c|}{1.6 ms \cite{1994-Equall-PRL}} &  & 164 $\mu$s \cite{1995-Equall-PRB} & 222 $\mu$s \cite{1995-Equall-PRB} \\ \hline  
        T$_{2,opt}$&  \multicolumn{2}{p{4cm}}{1.5 ms (B = 0), 2.6 ms (B = 10mT) \cite{1994-Equall-PRL}}&  \multicolumn{2}{|p{4.0cm}|}{1.1 ms (B = 0), 1.9 ms (B = 10 mT) \cite{1994-Equall-PRL}}&  &  152 $\mu$s (B = 7.7 mT) \cite{1995-Equall-PRB} &377 $\mu$s (B = 7.7 mT) \cite{1995-Equall-PRB} \\ \hline  
         T$_{1,HF}$&  \multicolumn{2}{|c|}{$>$ 20 days \cite{2003-Konz-PRB}}&  \multicolumn{2}{|c|}{$>$ 20 days \cite{2003-Konz-PRB}}&  & $\approx$ 100 s \cite{Nilsson2004} & $\approx$ 100 s \cite{Nilsson2004} \\ \hline 
         T$_{2,HF}$& 6 hours (B=1.37 T) \cite{2015-Zhong-Nature} &  &  &  &  & 0.5 ms (B=0) \cite{Ham1997},    42 s (B=80 mT) \cite{2013-Heinze-PRL} & 2.6 ms (B=0) \cite{Xiao2020}\\ \hline 
    \end{tabular}
\end{table*}


Kramers ions have lager  splitting of the ground doublet in an external magnetic field.
It was recently discovered that the dysprosium ion in a SrY$_{2}$O$_{4}$ crystal has a long electron spin magnetization lifetime ($\sim$1400 s) of the ground doublet at a temperature of 2 K \cite{2024-Malkin-PhysRevB}.\
This will allow creating  the spin polarization of the ground doublet  closer to 100\%.
The frequency of the echo signal and laser $\pi$-pulses can be greatly detuned from each other by selecting the required magnetic field.
 This will not require an additional filter crystal  for storage time $<$10$\cdot$T$_{1opt}$.
The T$_{2HF}$ and T$_{2opt}$ time values in this crystal are not yet known and require additional researches.

At ultralow temperatures ($<$50 mK), spin sublevels of other Kramers ions can be used, for example, in the optical ground state of Er$^{3+}$ in CaWO$_4$, where a spin coherence time of 23 ms was recently demonstrated \cite{2021-SciAdv-23ms}.
When a doublet is split in a magnetic field by $\approx$ 8 GHz at a temperature below 50 mK, almost $\approx 100\%$ spin polarization is achieved  automatically.
In low magnetic fields ($\sim$100 mT), the optical absorption lines of Er$^{3+}$ have a width of $\sim$1.75 GHz in the region of 1.5 $\mu$m.
The lines corresponding to the transitions between the two doublets are well spectrally resolved \cite{2024-OptoRfCaWO4} 
and can be used in the proposed QM,
which can be realized at higher temperatures (T $\sim$ 2K) in a Y$_2$SiO$_5$ crystal doped with $^{167}$Er$^{3+}$ ions.
It was shown \cite{Rancic2017} that in a strong magnetic field (B $>$ 3 T) the coherence times of the hyperfine sublevels and the optical transition become very large T$_{2,HF}$=1.3 s and T$_{2,opt}$=1.35 ms.
However, a preliminary modification of the optical absorption spectrum is required \cite{2021-Stuart-PRR}.
In the case of inhomogeneously broadened spin transition, it will be possible controlling spin PLM-coherence by applying the additional RF-pulses.
This will make it possible to repeatedly restore the spin PLM-coherence at different moments of time, allowing the retrieval of stored light pulses in the time-bin quantum states \cite{Moiseev2004c}.
Such retrieval schemes provides new opportunities of the proposed QM which deserve independent study.

\textit{Conclusion.-}
In this work, we propose an optical QM based on spin PLM-coherence, which, as a new quantum resource, manifests itself in the appearance of previously non-existent possibilities for the implementation of optical QM.
The use of spin PLM-coherence leads to the new physical properties of optical QM protocols, 
that significantly expand the functionality of QM and improve its basic parameters.
Our analysis also shows the existence of real experimental possibilities for the implementation of the proposed QM, the emergence of which will become a new chapter in the development of QM on atomic ensembles.
What other variants are possible for creating PLM-coherence and what new opportunities this will create for QM are also interesting topics for future research, especially in the case of using multi-level atomic ensembles.

 This research was supported by the Ministry of Science and Higher Education of the Russian Federation (Reg. number NIOKTR 121020400113-1).

\bibliography{apssamp.bib} 

\end{document}